\documentclass[acus]{JAC2000}

\usepackage{graphicx}

\begin{document}
\title{\flushright{WECT003}\\[15pt]
\centering READOUT AND CONTROL OF A POWER-RECYCLED \\
INTERFEROMETRIC GRAVITATIONAL WAVE ANTENNA
\thanks{Work supported
by National Science Foundation cooperative agreement PHY-9210038}}

\author{Daniel Sigg and Haisheng Rong, LIGO Hanford
Observatory, \\ P.O. Box 1970 S9-02, Richland, WA 99352 \\ Peter
Fritschel, Michael Zucker, Department of Physics and Center for
Space Research, \\ Massachusetts Institute of Technology,
Cambridge, MA 02139 \\ Rolf Bork, Nergis Mavalvala, Dale Ouimette,
LIGO Project, \\ California Institute of Technology, Pasadena, CA
91125 \\ Gabriela Gonz\'alez, Department of Physics and Astronomy,
Louisiana State University, \\ Baton Rouge,LA 70803}

\maketitle

\begin{abstract}
Interferometric gravitational wave antennas are based on Michelson
interferometers whose sensitivity to small differential length
changes has been enhanced by adding multiple coupled optical
resonators. The use of optical cavities is essential for reaching
the required sensitivity, but sets challenges for the control
system which must maintain the cavities near resonance. The goal
for the strain sensitivity of the Laser Interferometer
Gravitational-wave Observatory (LIGO) is $10^{-21}$~rms,
integrated over a $100$~Hz bandwidth centered at $150$~Hz. We
present the major design features of the LIGO length and frequency
sensing and control system which will hold the differential length
to within $5\times 10^{-14}$~m of the operating point. We also
highlight the restrictions imposed by couplings of noise into the
gravitational wave readout signal and the required immunity
against them.
\end{abstract}

\section{INTRODUCTION}

The interferometric gravitational wave detectors currently under
construction by LIGO\cite{ifos}, VIRGO\cite{ifo2}, GEO\cite{ifo3}
and TAMA\cite{ifo4} are expected to reach strain sensitivity
levels of $\sim\! 10^{-22} /\surd {\rm Hz}$ at 150 Hz over
baselines of several hundred meters up to several
kilometers\cite{weiss}. To achieve this sensitivity all of these
interferometers implement a Michelson laser interferometer
enhanced by multiple coupled optical
resonators\cite{meers,meers2}.

\begin{figure*}[t]
\centering
\includegraphics*[width=150mm]{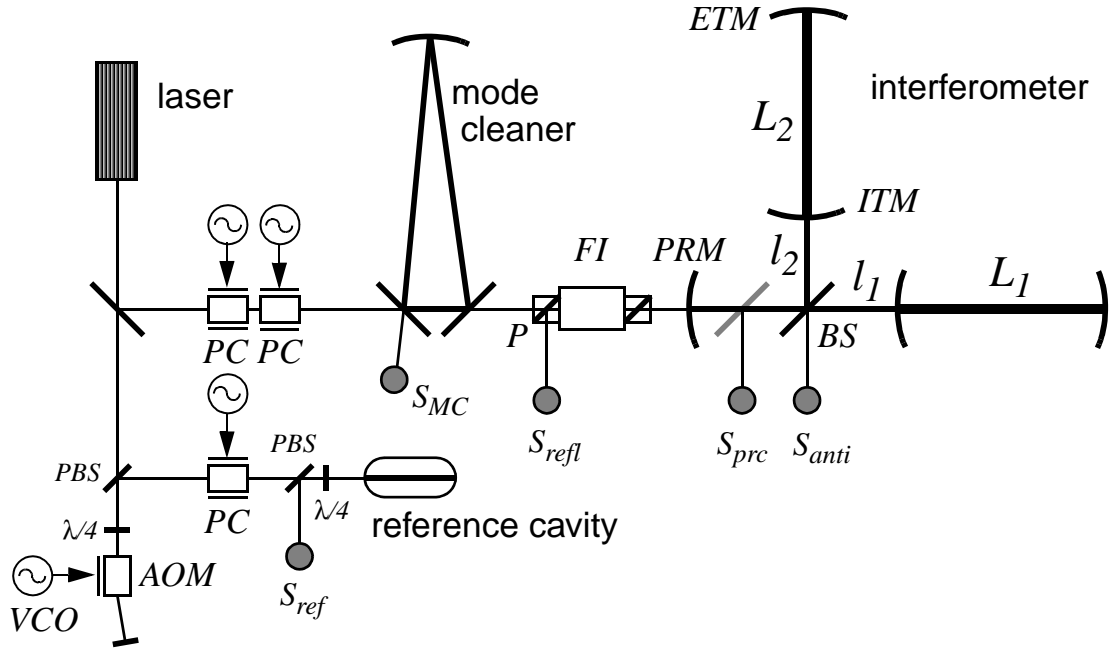}
\caption{Schematic view of the optical path in LIGO. The light of
a frequency stabilized Nd:YAG laser is passed through a triangular
mode cleaner cavity before it is launched into a Michelson
interferometer. To stabilize the laser frequency a small fraction
of the light is sampled, doubly passed through an acousto-optic
modulator (AOM) which serves as a frequency shifter, passed
through a Pockels cell and sent to a reference cavity. Using a
polarizing beamsplitter (PBS) and quarter-wave plate ($\lambda$/4)
the light reflected from the reference cavity is measured by a
photodetector to obtain the error signal, $S_{\rm ref}$, which in
turn is used to adjust the laser frequency. The main laser light
is passed through a pre-modecleaner (not shown) and two Pockels
cells which impose the phase-modulated rf sidebands used to lock
the mode cleaner and the Michelson interferometer. The mode
cleaner locking signal, $S_{\rm MC}$, is measured by a
photodetector in reflection of the mode cleaner cavity. The light
which passes through the mode cleaner is sent through a Faraday
isolator (FI) which also serves the purpose---together with a
polarizer (P)---to separate out the reflected light signal,
$S_{\rm refl}$. The main interferometer consists of a beamsplitter
(BS), two arm cavities each of them formed by an input test mass
(ITM) and an end test mass (ETM), and the power recycling mirror
(PRM). Additional locking signals are obtained at the
antisymmetric port, $S_{\rm anti}$, and by sampling a small amount
of light from inside the power recycling cavity, $S_{\rm prc}$.}
\label{fig1}
\end{figure*}

LIGO implements a power-recycled Michelson interferometer with
Fabry-Perot arm cavities (see Fig.~\ref{fig1}). Using optical
cavities is essential in reaching the ultimate sensitivity goal
but it requires an active electronic feedback system to keep them
``on resonance''. The control system must keep the round-trip
length of a cavity near an integer multiple of the laser
wavelength so that light newly introduced into the cavity
interferes constructively with light from previous round-trips.
Under these conditions the light inside the cavity builds up and
the cavity is said to be on resonance\cite{siegman}. Attaining
high power buildup in the arm cavities also requires that minimal
light is allowed to leave the system through the antisymmetric
port, so that all the light is sent back in the direction of the
laser where it is reflected back into the system by the power
recycling mirror. Hence, an additional feedback loop is needed to
control the Michelson phase so that the antisymmetric port is set
on a dark fringe.

\section{Environmental Influences}

It is important to distinguish low ($<50$~Hz) and high frequency
behaviour of the instrument. The low frequency region is typically
dominated by environmental influences many orders of magnitude
larger than the designed sensitivity and in many cases also many
orders of magnitude larger than what can be tolerated for stable
operations. It is the high frequency regime which yields good
sensitivity and which is used for detecting gravitational waves.
To suppress low frequency disturbances many active feedback
control systems are needed to compensate 4 longitudinal\cite{lsc}
and 14 angular\cite{asc} degrees-of-freedom in the main
interferometer alone. Additional feedback compensation networks
are needed to locally damp the suspended mirrors ($13 \times 4$
dofs), to control the mode cleaner (5 dofs) and to control the
laser (2 dofs).

For example, seismic motion of the ground\cite{seis} is many
orders of magnitude larger than the required gravitational wave
sensitivity. In LIGO a multi-stage passive seismic isolation
stack\cite{stack} together with a single-stage pendulum suspension
system\cite{sus} is used to isolate the optical components from
ground vibrations. This system system works well for frequencies
above $\sim\! 10 {\rm\,\, Hz}$, but gives no suppression at
frequencies of a Hz and below.

\section{Feedback Compensation Network}

In order to implement feedback each degree-of-freedom which is
under control of the compensation network has to be measurable.
LIGO implements the Pound-Drever-Hall reflection locking
technique\cite{pdh} to keep cavities on resonance and a variant of
this technique is used to control the angular
degrees-of-freedom\cite{align}. These techniques work well near
resonance where they behave linearly but have a strong non-linear
behaviour far way from resonance giving no or misleading signals.
The first step of engaging the feedback compensation network is to
catch the system on resonance with a highly sophisticated computer
code\cite{lock} running on a digital controls system.

\begin{figure}[t]
\centering
\includegraphics*[width=80mm]{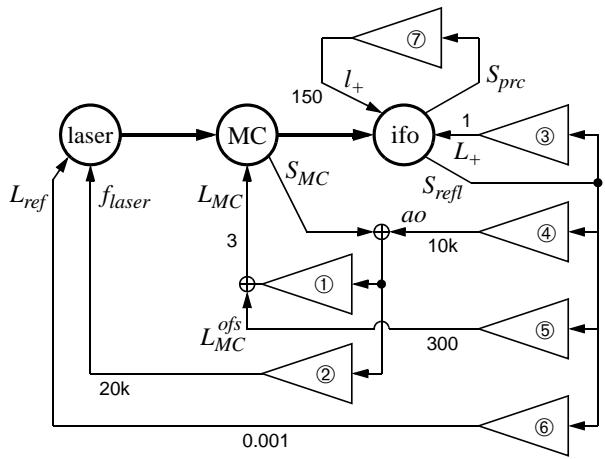}
\caption{Common mode control system. The mode cleaner error
signal, $S_{\rm MC}$, is split into two paths: the mode cleaner
length path~(1) feeding back to the position of a mode cleaner
mirror, $L_{MC}$, and the laser path~(2) feeding back to the laser
frequency, $f_{\rm laser}$, using the VCO/AOM. The in-phase
reflection signal, $S_{\rm refl}$, of the interferometer (ifo) is
split into four paths: the arm cavity path~(3) feeding back to the
common arm cavity mirror positions, $L_{+}$, the additive offset
(ao) path~(4) feeding back to the error point of the mode cleaner
control system, the mode cleaner length offset path~(5) feeding
back to the mode cleaner mirror position, $L_{MC}^{ofs}$, and the
tidal path~(6) feeding back to the reference cavity length,
$L_{\rm ref}$, using the thermal actuator. The in-phase signal at
the power recycling cavity port, $S_{\rm prc}$, is mostly
sensitive to the power recycling cavity length, $l_{+}$, and is
feed back to the recycling mirror position~(7). The numbers in the
feedback paths indicate unity gain frequencies in hertz.}
\label{fig2}
\end{figure}

A schematic view of the length control system for the common mode
degrees-of-freedom is shown in Fig.~\ref{fig2}. The signal $S_{\rm
refl}$ measuring the common arm length of the interferometer is
fed back to a combination of test masses, mode cleaner length and
laser frequency to achieve the required laser frequency noise
suppression of $<10^{-6}{\rm Hz}/\sqrt{\rm Hz}$ in the frequency
band of interest. To maintain maximum optical power in the
system---and thus maximum signal to shot noise ratio---the control
system must hold the common cavity length within
$<2\times10^{-12}{\rm\,\, m\,rms}$ of its resonance point. A
similar but less complicated system is deployed to control the
differential degrees-of-freedom. Their the differential arm cavity
length has to be held within $<5\times10^{-14}{\rm\,\, m\,rms}$ of
its operating point to not pollute the gravitational wave signal
with laser frequency noise.

\section{Conclusions}

So far LIGO has successfully demonstrated that the interferometer
can be locked and kept on resonance for hours. The main goal in
the near term is to improve the sensitivity which is still many
orders of magnitude away from design, to engage the remaining
feedback control paths and to fine-tune servo parameters.


\begin{thebibliography}{9}

\bibitem{ifos}A.~Abramovici, W.~Althouse, J.~Camp,
J.A.~Giaime, A.~Gillespie, S.~Kawamura, A.~Kuhnert, T.~Lyons,
F.J.~Raab, R.L.~Savage Jr., D.~Shoemaker, L.~Sievers, R.~Spero,
R.~Vogt, R.~Weiss, S.~Whitcomb, and M.~Zucker, ``Improved
sensitivity in a gravitational wave interferometer and
implications for LIGO,'' Phys. Lett. {\bf A218}, 157--163 (1996).

\bibitem{ifo2}B.~Caron, A.~Dominjon, C.~Drezen, R.~Flaminio,
X.~Grave, F.~Marion, L.~Massonnet, C.~Mehmel, R.~Morand, B.~Mours,
V.~Sannibale, M.~Yvert, D.~Babusci, S.~Bellucci, S.~Candusso,
G.~Giordano, G.~Matone, J.-M.~Mackowski, L.~Pinard, F.~Barone,
E.~Calloni, L.~DiFiore, M.~Flagiello, F.~Garuti, A.~Grado,
M.~Longo, M.~Lops, S.~Marano, L.~Milano, S.~Solimeno, V.~Brisson,
F.~Cavalier, M.~Davier, P.~Hello, P.~Heusse, P.~Mann, Y.~Acker,
M.~Barsuglia, B.~Bhawal, F.~Bondu, A.~Brillet, H.~Heitmann,
J.-M.~Innocent, L.~Latrach, C.N.~Man, M.~PhamTu, E.~Tournier,
M.~Taubmann, J.-Y.~Vinet, C.~Boccara, P.~Gleyzes, V.~Loriette,
J.-P.~Roger, G.~Cagnoli, L.~Gammaitoni, J.~Kovalik, F.~Marchesoni,
M.~Punturo, M.~Beccaria, M.~Bernardini, E.~Bougleux, S.~Braccini,
C.~Bradaschia, G.~Cella, A.~Ciampa, E.~Cuoco, G.~Curci,
R.~DelFabbro, R.~DeSalvo, A.~DiVirgilio, D.~Enard, I.~Ferrante,
F.~Fidecaro, A.~Giassi, A.~Giazotto, L.~Holloway, P.~LaPenna,
G.~Losurdo, S.~Mancini, M.~Mazzoni, F.~Palla, H.-B.~Pan,
D.~Passuello, P.~Pelfer, R.~Poggiani, R.~Stanga, A.~Vicere,
Z.~Zhang, V.~Ferrari, E.~Majorana, P.~Puppo, P.~Rapagnani, and
F.~Ricci, ``The VIRGO interferometer for gravitational wave
detection,'' Nucl. Phys. {\bf B54}, 167--175 (1997).

\bibitem{ifo3}K.~Danzmann, ``GEO 600 --- A 600-m Laser
Interferometric Gravitational Wave Antenna,'' in {\it First
Edoardo Amaldi conference on gravitational wave experiments},
E.~Coccia, G.~Pizella and F.~Ronga, eds. (World Scientific,
Singapore, 1995), p. 100--111.

\bibitem{ifo4}K.~Tsubono, ``300-m Laser Interferometer
Gravitational Wave Detector (TAMA300) in Japan,'' in {\it First
Edoardo Amaldi conference on gravitational wave experiments},
E.~Coccia, G.~Pizella and F.~Ronga, eds. (World Scientific,
Singapore, 1995), p. 112--114.

\bibitem{weiss} R.~Weiss, ``Electromagnetically coupled broadband
gravitational antennae,'' MIT Res. Lab. Electron. Q. Prog. Rep.
{\bf 105}, 54--76 (1972).

\bibitem{meers}J.-Y.~Vinet, B.J.~Meers, C.N.~Man, and A.~Brillet,
``Optimization of long-baseline optical interferometers for
gravitational-wave detection,'' Phys. Rev. {\bf D38}, 433--447
(1988).

\bibitem{meers2} B.J.~Meers, ``The frequency response of
interferometric gravitational wave detectors,'' Phys. Lett. {\bf
A142}, 465--470 (1989).

\bibitem{siegman}A.E.~Siegman, {\it Lasers,} (University Science,
Mill Valley, Calif., 1986), Chap.~13, p.~663.

\bibitem{lsc}P.~Fritschel, R.~Bork, G.~Gonz\'alez, N.~Mavalvala,
D.~Ouimette, H.~Rong, D.~Sigg, and M.~Zucker, ``Readout and
control of a power-recycled interferometric gravitational wave
antenna,'' Appl. Opt. {\bf 40}, 4988--4998 (2001).

\bibitem{asc}P.~Fritschel, G.~Gonz\'alez, N.~Mavalvala,
D.~Shoemaker, D.~Sigg, and M.~Zucker, ``Alignment of a
long-baseline gravitational wave interferometer,'' Appl. Opt. {\bf
37}, 6734--6747 (1998).

\bibitem{stack} J.~Giaime, P.~Saha, D.~Shoemaker, and L.~Sievers,
``A passive vibration isolation stack for LIGO: design, modeling,
and testing,'' Rev. Sci. Instrum. {\bf 67}, 208--214 (1996).

\bibitem{sus} A.~Gillespie and F.~Raab, ``Thermal noise in the
test mass suspensions of a laser interferometer gravitational-wave
detector prototype,'' Phy. Lett. {\bf A178}, 357--363 (1993).

\bibitem{seis} T.~Lay, and T.C.~Wallace, {\it Modern global
seimology,} (Academic Press, San Diego, California, 1995), p.~179.

\bibitem{pdh} R.W.P.~Drever, J.L.~Hall, F.V.~Kowalski, J.~Hough,
G.M.~Ford, A.J.~Munley, and H.~Ward, ``Laser phase and frequency
stabilization using an optical resonator,'' Appl. Phys. {\bf B31},
97--105 (1983).

\bibitem{align}
Y.~Hefetz, N.~Mavalvala, and D.~Sigg, ``Principles of calculating
alignment signals in complex optical interferometers,'' J. Opt.
Soc. Am. {\bf B14}, 1597--1605 (1997). \\ D.~Sigg, and
N.~Mavalvala, ``Principles of calculating the dynamical response
of misaligned complex resonant optical interferometers,'' J. Opt.
Soc. Am. {\bf A17}, 1642--1649 (2000).

\bibitem{lock}
M.~Evans, N.~Mavalvala, P.~Fritschel, R.~Bork, R.~Gustafson,
W.~Kells, M.~Landry, D.~Sigg, R.~Weiss, S.~Whitcomb, and
H.~Yamamoto, ``Lock acqusition of a gravitational wave
interferometer,'' submitted to Opt. Lett. (2001).

\end{thebibliography}
\end{document}